\documentclass[useAMS,usenatbib]{mn2e}
\usepackage{graphicx}
\usepackage{yfonts}
\usepackage{times}
\title[Photometric study of TV~Boo]{Photometric study of the star with changing Blazhko effect: TV~Bootis}
\author[M. Skarka and M. Zejda]{M. Skarka$^{1}$\thanks{E-mail: maska@physics.muni.cz (MS)} and M. Zejda$^{1}$\thanks{E-mail: zejda@physics.muni.cz (MZ)}
\\
$^{1}$Department of Theoretical Physics and Astrophysics, Faculty of Science, Masaryk University, Brno 61137, Czech Republic\\}
\begin{document}

%\date{Accepted 1988 December 15. Received 1988 December 14; in original form 1988 October 11}

\pagerange{\pageref{firstpage}--\pageref{lastpage}} \pubyear{2012}

\maketitle

\label{firstpage}

\begin{abstract}
This study investigates periodic modulation of the light curve of the RRc star TV Boo and 
its physical parameters based on photometric data. This phenomenon, known as Blazhko effect, 
is quite rare among RRc stars. Frequency analysis based on the data gathered at Masaryk 
University Observatory (MUO) and also using SuperWASP data revealed symmetrically 
structured peaks around the main pulsation frequency and its harmonics, which indicate 
two modulation components of the Blazhko period. The main modulation periodicity was found to 
be $9.7374\pm0.0054$~d. This is one of the shortest known Blazhko periods among RRc stars. 
The second modulation period ($21.5\pm0.2$~d) causes changes of the Blazhko effect itself. 
Some indices show that TV Boo could be affected by long-term changes in the order of years. 
Basic physical parameters were estimated via a MUO $V$ light curve solution using Fourier parameters. 
TV Boo seems to be a low-metalicity star with $\mathrm{[Fe/H]}=-1.89$.

\end{abstract}

\begin{keywords}
stars: horizontal branch -- stars: individual (TV Bootis)
\end{keywords}

\section{Introduction}

Light curve modulation together with cyclic period changing is quite common among
pulsation variables. In the case of RR Lyrae type stars such behaviour has been known for
more than one century and it is called the Blazhko effect after S. N. Blazhko, who
firstly noticed phase modulation in RW Dra \citep{blazhko1907}. The
amplitude modulation was discovered several years later by \citet{shapley} in the 
prototype RR Lyr itself. Nowadays more than 40 \% of all RRab stars seem to show this effect
\citep{kolenberg2010a,jurcsik2006a}, but only about 5~\% of RRc type
\citep{moskalik2002}.

Although many theories have been suggested to explain the observed properties of
modulated stars, the origin of the Blazhko effect is still not well understood. A brief
overview of these theories can be found in \citet{smith1995}, \citet{kovacs2009}  and
\citet{kolenberg2011a}. In the era of the {\it Kepler\/} space telescope the 9:2
resonance between the fundamental radial mode and the 9th overtone seems to be
responsible for the Blazhko effect \citep{buchler2011}.

The Blazhko effect manifests itself as sidelobe frequencies around the main pulsation
frequency and its harmonics in the frequency spectra. Also the modulation frequency 
itself and its harmonics are natural manifestations of the modulation (e.g. \citet{benko2011}, 
\citet{szeidl2009}). Frequencies are usually equally spaced and form doublets, triplets or 
higher multiplets.

The variability of TV Boo ($\alpha=14^{\mathrm{h}}16^{\mathrm{m}}36.58^{\mathrm{s}}$,
$\delta=+42^{\circ}21'35.69''$, J2000) was first reported in
\citet{gutnick1926}.GCVS \citep{samus2012} gives the period $P=0.3125609$ d and spectral type A7--F2.
Metallicity ($\mathrm{[Fe/H]}=-2.44$) listed in \citet{feast2008} makes TV Boo one of the
stars with the lowest metallicity among RR Lyrae type stars.

The Blazhko effect in TV Boo has been known since 1965, when \citet{detre1965} found a
Blazhko period of 33.5 d. Almost ten years later \citet{firmanyuk1974} reported the
Blazhko effect with a period of $16.14$ d. More recently, using data of the Northern Sky
Variability Survey (NSVS, \citet{wozniak2004}), \citet{wils2006} found the Blazhko period
to be 10 d.

\section{Observation and data reduction}

TV Bootis was observed on 23 nights during three seasons from 2009 to 2011 (for the complete observation 
log see tab. 1) with the 60cm Newtonian telescope of Masaryk University Observatory (MUO) in Brno, Czech 
Republic. This telescope is equipped with a ST-8 CCD with a KAF-1600 chip (FOV $17.0'\times11.3'$). Observations 
were carried out in $BV(R_{c})$ filters. In each of these passbands more than 2000 data points were collected.
\begin{table*}
 \centering
 \begin{minipage}{120mm}
  \caption{Observation log}
  \begin{tabular}{@{}lcccccc@{}}
  \hline
    Season & Time-span  & Number of nights & Number of hours & \multicolumn{3}{c}{Number of data points}\\
             & $(JD-2450000)$ &                             &    &  $B$ &     $V$  & $(R_{c})$ \\ \hline
    2009 & $4921-4971$      & 8                         &    50.9 & 1125 &    1173  & 1228 \\
    2010 & $5287-5377$      & 3                         &    9.5 &   -  &    307     & 356\\
    2011 & $5614-5683$    & 12                      &    53.4&  956 &    1180  & 1190\\
    Total&                              & 23                        &    113.8& 2081 &       2660  & 2774\\
\hline
\end{tabular}
\end{minipage}
\end{table*}
The data reduction and differential photometry were performed using the C-Munipack 
package\footnote{http://c-munipack.sourceforge.net/}. TYC 3038-955-1 and TYC 3038-1064-1 were used as 
comparison and check stars respectively. Although both stars are significantly redder than TV Boo itself 
(which could slightly affect the light curve and mainly color indices), these stars were chosen because 
of the lack of other suitable stars in the vicinity of TV Boo. Standard deviations in the difference 
between the comparison and check stars were better than 0.01 mag in all pasbands. All the points were 
transformed to the standard $BV(R_{c})$ magnitudes using standard stars in Landolt fields \citep{landolt1992}. 
MUO $V$ data are in fig.~1. The first two lines of these data are listed in tab. 2, the whole dataset is 
available as supporting information with the online version of this paper.

We also analysed SuperWASP data. These data are of better quality for frequency analysis than MUO data because of the
number of points (8914 points in 205 nights) and their time-span (1138~d). 

        \begin{figure}%[htbp]
            \begin{center}
            \includegraphics[scale=0.35]{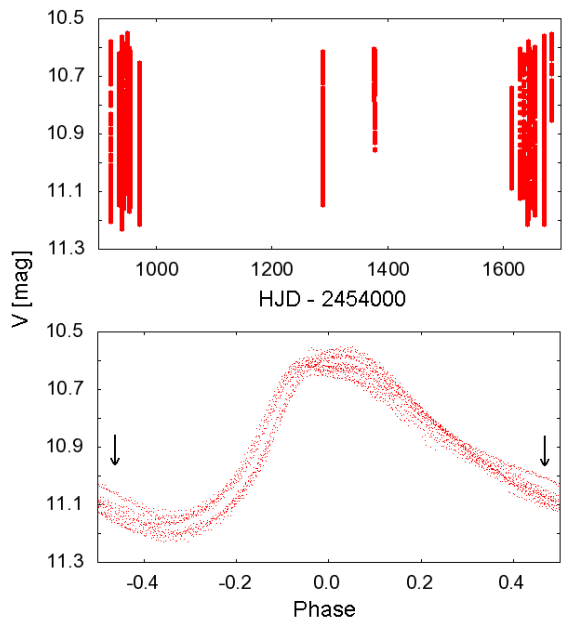}
            \caption{Standard $V$ data gathered on MUO between seasons 2009 and 2011. In the bottom panel the 
            data are phased according to eq.1. The hump on the descending branch is marked by an arrow.}
            \end{center}
        \end{figure}

        \begin{figure}%[htbp]
            \begin{center}
            \includegraphics[scale=0.24]{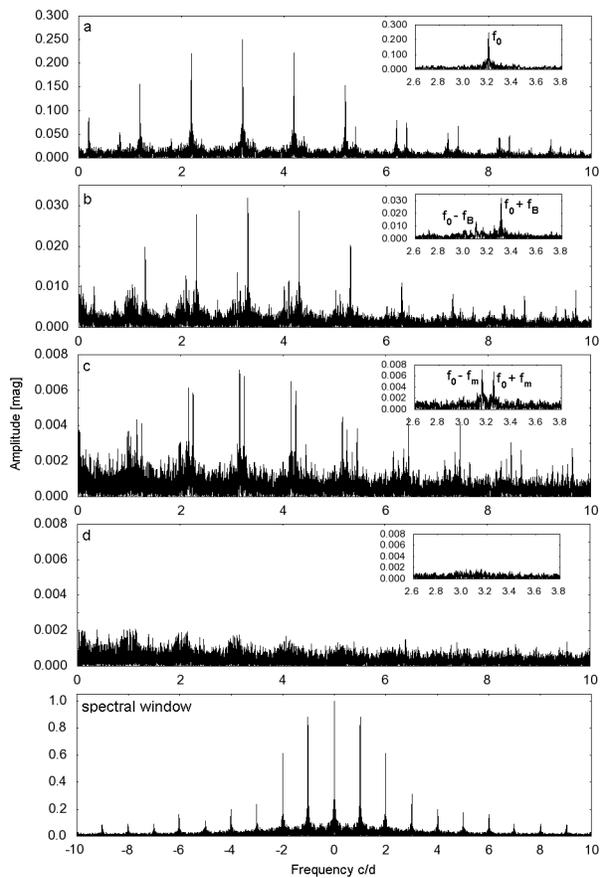}
            \caption{Frequency spectra of TV Boo based on SuperWASP data. 
            The top panel shows the Fourier spectrum with the main pulsation frequency (the 
            highest peak). Other noticeable peaks are daily aliases of the main pulsation frequency. 
            The next two panels show prewhitened spectra after removal of the main pulsation 
            frequency and its harmonics (panel b), after removal of the frequencies related to 
            the Blazhko effect (panel c) and final residual spectra with all modulation frequencies 
            removed (panel d). The vicinity of the main pulsation frequency with frequency 
            identifications are plotted in subplots of panels (a) to (d). The bottom panel shows 
            the spectral window.}
            \end{center}
        \end{figure}

\begin{table}
	\centering
  \caption{Muo $V$ data}
  \begin{tabular}{lr}
  \hline
JDhel   & $V$[mag] \\	\hline
2454921.515	&	11.201	\\
2454921.516	&	11.204	\\
\dots       & \dots   \\
\hline
	\end{tabular}
	\end{table}

\section{Frequency analysis}

\begin{table*}
 \centering
 \begin{minipage}{100mm}
  \caption{Frequencies found in MUO and SuperWASP data sets}
  \begin{tabular}{@{}cccccccccc@{}}
  \hline
                    &\multicolumn{3}{c}{MUO} & \multicolumn{3}{c}{SuperWASP}  \\
    Id          & Frequency &   $A_{V}$ &   S/N & Frequency &   $A$ &   S/N   \\
                &   c/d     &     mag   &       &   c/d     &   mag &         \\ \hline
$ f_{0}$        &   3.1993740(8)&   0.2620(3)   &   163.9   &   3.1993646(4) &   0.2497(3)   &   326.0  \\        
$ f_{0}+f_{B}$  &   3.302036(6) &   0.0342(6)   &   21.5    &   3.302066(3)  &   0.0318(3)   &   42.0   \\        
$ f_{0}-f_{B}$  &   3.09671(2)  &   0.0109(4)   &   6.8     &   3.09667(1)   &   0.0093(3)   &   11.9   \\        
$ f_{0}+f_{m}$  &   3.244555(3) &   0.0066(6)   &   4.1     &   3.24598(2)   &   0.0060(3)   &   7.9    \\        
$ f_{0}-f_{m}$  &               &               &           &   3.15276(2)   &   0.0070(3)   &   9.1    \\        
$ 2f_{0}$       &   6.398748(4) &   0.0730(5)   &   49.1    &   6.398738(2)  &   0.0729(3)   &   156.6  \\        
$2f_{0}+f_{B}$  &   6.50141(2)  &   0.0133(5)   &   9.0     &   6.50143(2)   &   0.0092(3)   &   19.9   \\        
$2f_{0}-f_{B}$  &   6.29609(3)  &   0.0072(5)   &   4.8     &   6.29603(2)   &   0.0069(3)   &   14.8   \\        
$2f_{0}+f_{m}$  &               &               &           &   6.44534(3)   &   0.0040(4)   &   8.7    \\        
$2f_{0}-f_{m}$  &               &               &           &   6.35212(4)   &   0.0026(3)   &   5.6    \\        
$ 3f_{0}$       &   9.59812(1)  &   0.0187(4)   &   14.8    &   9.598107(6)  &   0.0185(3)   &   43.6   \\        
$3f_{0}+f_{B}$  &   9.70078(3)  &   0.0102(4)   &   5.7     &   9.70080(1)   &   0.0077(3)   &   18.2   \\        
$3f_{0}-f_{B}$  &               &               &           &   9.49540(3)   &   0.0034(3)   &   8.0    \\        
$3f_{0}+f_{m}$  &               &               &           &   9.64471(5)   &   0.0024(3)   &   5.6    \\        
$ 4f_{0}$       &   12.79750(2) &   0.0088(4)   &   9.1     &   12.79748(1)  &   0.0107(3)   &   29.3   \\        
$4f_{0}+f_{B}$  &   12.90016(4) &   0.0051(4)   &   5.2     &   12.90020(3)  &   0.0032(3)   &   8.7    \\        
$4f_{0}-f_{B}$  &               &               &           &   12.69480(6)  &   0.0021(3)   &   5.7    \\        
$ 5f_{0}$       &   15.99687(2) &   0.0090(4)   &   11.7    &   15.99685(1)  &   0.0079(3)   &   22.5   \\        
$5f_{0}+f_{B}$  &               &               &           &   16.09950(6)  &   0.0019(3)   &   5.6    \\        
$5f_{0}+f_{m}$  &               &               &           &   16.04341(7)  &   0.0016(3)   &   4.6    \\        
$ 6f_{0}$       &   19.19624(4) &   0.0052(4)   &   7.7     &   19.19621(2)  &   0.0050(3)   &   15.4   \\        
$ 7f_{0}$       &   22.39562(5) &   0.0037(4)   &   6.4     &   22.39558(3)  &   0.0032(3)   &   11.2   \\        
$ 8f_{0}$       &   25.59499(5) &   0.0018(4)   &   4.6     &   25.59495(6)  &   0.0019(3)   &   6.9    \\\hline  
\end{tabular}
\end{minipage}
\end{table*}

Frequency analysis was performed using the MUO data set as well as data from the SuperWASP \citep{butters2010} 
and NSVS \citep{wozniak2004} surveys. As the main tool for frequency analysis Period\,04 \citep{lenz2004} was 
used. For basic pulsation period determination PerSea software (made by G. Maciejewski on the basis of 
\citet{schwarzenberg1996} method) and a Matlab script based on nonlinear least square method were used. Within 
their margins of error both methods gave the same results, so the maxima timings can be expressed as:

\begin{equation}
   \mathrm{HJD}~T_{\mathrm{max}}=2454922.2334(2)+0.3125615(7)E_{\mathrm{puls}}.
\end{equation}
        
Analysis based on the MUO dataset gave almost the same results as were obtained with the SuperWASP data. 
Detected frequencies in all studied datasets are listed in tab. 3. Semi-amplitudes, S/N
ratios\footnote{We mention only peaks with $S/N>4$, because this is a generally accepted limit 
to distinguish between frequency peaks due to pulsation and noise \citep{breger1993}.} 
for individual frequencies and their standard deviations from the results obtained from Monte Carlo 
simulations are also given in this table. Results for NSVS data are not listed, because only three 
frequencies ($f_{0}$, $2f_{0}$ and $f_{0}+f_{B}$) were detected in this dataset.

Frequency analysis showed two modulation components located symmetrically around the main 
pulsation frequency and its harmonics $kf_{0}$ up to the third order in the case of 
SuperWASP data and only one component related to the second modulation frequency ($f_{0}+f_{m}=3.244555$~c/d) 
was revealed in MUO data set. We detected 8 harmonics of the main pulsation frequency in all studied datasets.

The positions of sidelobe peaks with higher amplitudes $(kf_{0}\pm f_{B})$ correspond to 
$f_B=0.102697$ cycles/day, which results in a 9.7374~d modulation period (Blazhko period). 
Except for the sidelobe frequencies, other equally spaced peaks were detected in SuperWASP 
data around the main frequency with values $f_{0}+f_m=3.24598$ c/d and $f_{0}-f_m=3.15276$ c/d. The 
difference between these frequencies and the main pulsation frequency is $0.04661$ c/d, 
which indicates second modulation period of length $21.5$~d. This additional period 
is usualy explained as a changing Blazhko effect, which is also known in other stars, 
e.g. XZ Cyg \citep{lacluyze2004}, RV UMa \citep{hurta2008} or LS Her \citep{wils2008}. 
A similar frequency structure with separate modulation peak series was also revealed 
in the case of CZ Lac \citep{sodor2011} and V445 Lyr \citep{guggenberger2012}. All named stars 
(except LS Her) are of the RRab type. TV Boo is therefore the first known RRc type star 
with such a frequency structure\footnote{In the case of LS Her interesting frequency structure with an 
equidistant group of three frequencies on both sides of the main pulsation frequency was 
detected \citep{wils2008}}.

According to the frequency analysis the maxima timings for the Blazko effect can be expressed as
\begin{equation}
   \mathrm{HJD}~T_{\mathrm{maxBlazhko}}=2454950.80(10)+9.7374(54)E_{\mathrm{Blazhko}}.
\end{equation}
        
Phased MUO $V$ and SuperWasp light curves look very similar in shape and amplitude. Therefore, we 
tried to combine these datasets into one to evaluate the result. Phased light curves were fitted with 
the linear combination of sines to get a mean light curve and finally to obtain the shift in magnitudes between 
these two data sets\footnote{Zero points and amplitudes of the light changes were 11.123 mag 
$\Delta m$ 0.534 mag in the case of SuperWASP data and 10.917 mag and $\Delta m$ 0.567 mag for MUO $V$ data, 
respectively. The shift was therefore 0.206 mag.}. Frequency analysis of such dataset revealed five frequency 
peaks related to the second modulation frequency, which were not detected in both datasets when studied separately. 
SuperWasp data cover a much broader wavelength range than the Johnson V passband. In addition the shape of SuperWasp 
and MUO $V$ light curve can be slightly different, therefore such combination is not rigorous.
        
One of the recent highlights of RR Lyrae research is the discovery of period-doubling \citep{szabo2010} 
and high-order resonances \citep{kollath2011} as possible causes of the Blazhko effect. No signs of such behaviour 
were unveiled in the frequency spectra of TV~Boo. This could either be caused by insufficient quality of the data 
for this purpose or first overtone Blazhko pulsators do not show such behaviour.

Another puzzle seen in many RRc stars (and Cepheids) observed with high-precision photometry is the 
presence of an additional frequency with a strangely repeating frequency ratio of 1.58--1.63 from star to 
star (see e.g. \citet{moskalik2012}). Again, no frequency peaks with the given ratio or even close to this ratio 
were identified in studied datasets down to the level of residuals. In addition, the detection of such 
a frequency was complicated by the fact that in the area defined by the ratio fell aliases of the main pulsation 
frequency and also aliases related to modulation frequencies. In any case, based on our analysis, we could not 
confirm or disprove the presence of the frequency in the range of 1.58--1.63 $f_{0}$. The fact, that such a feature 
could not be confirmed might indicate that the feature is absent in the case of TV Boo.

Frequency spectra together with the spectral window resulting from the SuperWASP dataset 
are shown in fig.~2. The first two panels show Fourier spectra with the main pulsation 
frequency (a) and spectrum after removal of this frequency and
its harmonics (b). There are easily noticeable daily aliases of the highest peaks in
these two panels. The final residual spectrum, after prewhitening with all basic and
modulation frequencies, is given in panel (d). The height (about 3\,mmag) of the
residual spectrum is probably caused by small night-to-night shifts in the data
as well as by unequally spaced observations. There is no significant peak in the residual
plot.

        \begin{figure}%[htbp]
            \begin{center}
            \includegraphics[scale=0.35]{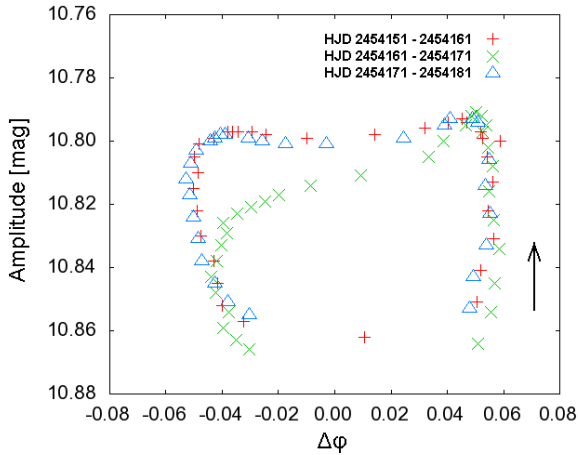}
            \caption{Phase-amplitude diagram in three Blazhko cycles between HJD 2454151 and 
            2454181. The run of points is counterclockwise similarly as in majority of Blazhko RRab stars.}
            \end{center}
        \end{figure}

Fig.~5 shows what the light curve looked like during one Blazhko cycle between HJD 2454151 
and 2454161. Each Blazhko cycle affects the light curve shape 
uniquely in a given Blazhko phase, which is the manifestation of a changing Blazhko effect. 
Notice the change of light curve shape around maxima during the Blazhko cycle.

% zalozena nova sekce pro popis vzhledu krivky a fazovych shiftu

\section{Remarks on the period change during Blazhko cycle and possible long term features}

Blazhko RRc stars primarily show frequency modulation with only small amplitude modulation. 
In TV Boo the amplitude of the light curve modulation is only about $0.1$ mag and the maximum and minimum of light 
changes are affected to a similar extent and the fuzziness of the light curve is uniform in all its 
parts (check fig.~1). 
        
TV Boo, as a frequency modulated star with only small amplitude modulation, shows 
changes in pulsation period during the Blazhko cycle. Changes in phase and amplitude of maximum 
light during three unique Blazhko cycles between HJD 2454151 and 2454181 can be seen in fig.~3. 
Points in this plot represent maxima of our light curve model. Points from the first and third Blazhko cycles 
create curves of almost the same shape, but the curve based on points coming from the cycle between 
them differs. The phase-amplitude diagram will be similar for a few following Blazhko cycles before 
it will change considerably. Bear in mind that the shape of the light curve is slowly changing in time, so 
the phase-amplitude diagram will look completely different after many Blazhko cycles. 

        \begin{figure}
            \begin{center}
            \includegraphics[scale=0.27]{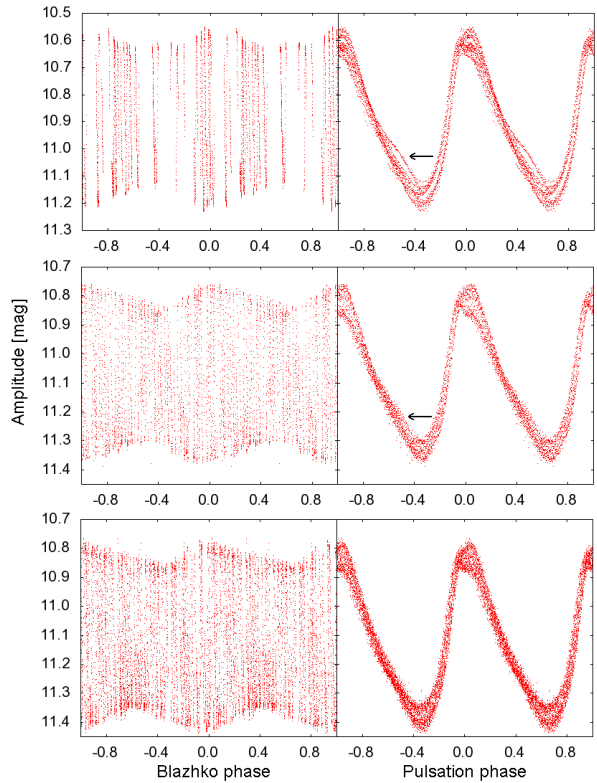}
            \caption{MUO $V$ data (top panel), SuperWASP data from season 2004 (middle panel) and 
            from season 2007 (bottom panel). Data folded with the Blazhko period are on the left panels of the picture,
						data phased with the main pulsation period are to the right. Note the hump on the descending branch of MUO 
						data and SuperWASP data (marked by an arrow).}
            \end{center}
        \end{figure}

The amplitude of the period changes determined from maxima timings is about 
$0.13 P_{\mathrm{puls}}$, in the terms of time it is about 58 minutes. This value is strongly affected 
by the change of the light curve shape around maxima. A more realistic estimation of the 
magnitude of the phase modulation can be obtained by measuring the phase shift in the ascending 
branch. We measured the phase shift in fixed magnitude 11.1 (approximately half of the 
light curve amplitude) and we obtained the strength of period change during Blazhko cycle 
beginning at HJD 2454151 shown in fig. 5 $0.046 P_{\mathrm{puls}}$, which gives a little less than 21 minutes. 
This value is also slightly cycle-to-cycle dependent, but only on the order of $10^{-3}P_{\mathrm{puls}}$.
        
     \begin{figure*}
         \centering
            \begin{minipage}{170mm}
            \includegraphics[scale=0.23]{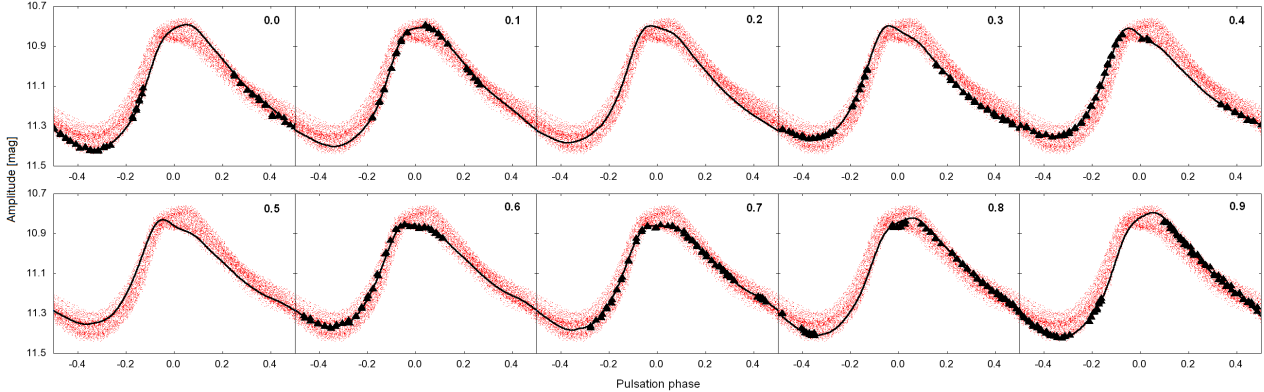}
            \caption{Light curve changes during one unique Blazhko cycle between HJD 2454151 and 2454161. 
            The modelled light curve belonging to the given Blazhko phase (mentioned in the top right 
            corner of each plot) is plotted with a continuous line. The triangles show real SuperWASP data 
            for the given Blazhko phase in this unique time-span. All available SuperWASP data (plotted with dots) 
            are just for illustration to see the range of possible light curves.}
            \end{minipage}
    \end{figure*}

Descending branches of the light curves are deformed with humps in some minima of the Blazhko cycles, but they are not 
present in each Blazhko cycle. In the MUO data the hump was not detected in the season 2009, but it is clearly visible 
in the data gathered in 2010 and 2011. Similarly, in the SuperWASP data from the season 2007 this behaviour was not 
unveiled, but there is a weak sign of the hump in the 2004 season (see fig.~4, middle panel). In the same figure it is apparent 
that the Blazhko light curve is strongly asymmetric with its minimum around the Blazhko phase 0.7. This feature could be a 
demonstration of long-term changes known in few other examples, e.g. in RR Lyrae itself with long changes lasting 4 
years \citep{detre1973}. A first order estimation of such a period could then be about 6 years. However this claim 
is based on the assumption that MUO $V$ and SuperWASP light curves are of the same shape and amplitudes. Of course, 
absence of the hump can be caused by the properties of the data itself - lack of data for such Blazhko phases in studied 
datasets or the overlapping of data by other data.

\section{Physical parameters}

In the 1990's some empirical relations between so-called Fourier coefficients and basic physical parameters 
of RR Lyraes were found, namely there are calibrations for metallicity, temperature, absolute magnitude etc. 
\citep{kovacs1998,jurcsik1998}. The method is based on the decomposition of a phased $V$ band light curve into 
its sum of sines or cosines. This can be done for stable RR Lyraes and even for Blazhko stars, but good coverage 
of the folded light curve is needed. After decomposition Fourier parameters $R_{ij}$ and $\phi_{ij}$ can be determined as follows:

\begin{equation}
V=A_{0}+\sum_{i=1}^n A_{i}\sin(i\omega t+\phi_{i}),
\end{equation}
\begin{equation}
\phi_{ij}=j\phi_{i}-i\phi_{j},
\end{equation}
\begin{equation}
R_{ij}=A_{i}/A_{j}.
\end{equation}

The phased MUO $V$ light curve was modelled with the sum of sine with 9 components (fig.~6). The order of fit 
was chosen by visual inspection. The first few obtained Fourier parameters are given in tab. 4.

The following calibrations for RRc stars listed in \citet{simon1993,morgan2007,kovacs1998} were used:

\[\left[\mathrm{Fe/H}\right]=52.466P^{2}-30.075P+0.131\left(\phi^{c}_{31}\right)^{2}\]
\begin{equation}
~~~~~~~~~+0.9825\phi^{c}_{31}-4.198\phi^{c}_{31}P+2.424
\end{equation}

\begin{equation}
M_{\mathrm{V}}=1.061-0.961P-4.447A_{4}-0.044\phi^{s}_{21}
\end{equation}

\begin{equation}
\log(L/\mathrm{L_{\odot}})=2.41+1.04 \log P-0.058\phi^{c}_{31}
\end{equation}

\begin{equation}
\log T_{\mathrm{eff}}=3.7746-0.1452\log P+0.0056\phi^{c}_{21}
\end{equation}

\begin{equation}
\log (M/\mathrm{M_{\odot}})=0.39+0.52\log P-0.11\phi^{c}_{31},
\end{equation}
where $\phi^{c}_{ij}$ are coefficients based on cosine-term decomposition and $\phi^{s}_{ij}$ 
sine-term decomposition coefficients respectively. The relations between sine and cosine 
coefficients are $\phi^{s}_{ij}=\phi^{c}_{ij}-(j-k)\frac{\pi}{2}$. The zero point in eq. 7 
was decreased according to \citet{cacciari2005} by 0.2 as opposed to the equation in \citet{kovacs1998}.

\begin{table}
 \centering
  \caption{Fourier parameters based on sine-term decomposition.}
  \begin{tabular}{@{}lcccccc@{}}
  \hline
    $A_{0}$ & $R_{21}$  & $R_{31}$  & $R_{41}$  & $\phi_{21}$   & $\phi_{31}$ & $\phi_{41}$ \\
        mag         &   mag             &   mag             &   mag             &   rad                     &   rad                 &   rad \\  \hline

        10.917              &   0.283   &   0.078   &   0.043   &   2.698   &   5.545   &   1.996   \\
\hline
\end{tabular}
\end{table}

\begin{table}
 \centering
  \caption{Mean physical parameters of TV Boo.}
  \begin{tabular}{@{}ll@{}}
  \hline
~~~$[\mathrm{Fe/H}]$~~~~~~~~~~~~~~~~~~~~~~~~~~~~~~~~~~~~    &   $-1.89\pm0.14$~~~   \\
~~~$T_{\mathrm{eff}}$ [K]   &   ~\,$7270\pm50$  \\
~~~$M_{\mathrm{V}}$ [mag]   &   ~~$0.59\pm0.06$ \\
~~~$L [\mathrm{L_{\odot}}]$&    ~~$56\pm5$  \\
~~~$R [\mathrm{R_{\odot}}]$& ~~$4.8\pm0.5$ \\
~~~\textfrak{M}\,[\textfrak{M}$_{\odot}]$   &   ~~$0.73\pm0.04$ \\
~~~$r$ [pc] &   ~\,$1150\pm30$  \\
\hline
\end{tabular}
\end{table}

Physical parameters resulting from eq. 6--10 using the MUO $V$ light curve fit noted in tab. 4, 
are listed in tab. 5. Metallicity in eq. 6 is given in the \citet{zinn1984} scale. The solution, 
and mainly parameter $\phi_{31}$, is very sensitive to the coverage of the light curve. If we 
remove three nights when the hump was observed, the $\phi_{31}$ decreases its value by about 0.04, 
which leads to a change in $[\mathrm{Fe/H}]$ of about $-0.4$. Our light curve is not ideally 
covered in all possible Blazhko phases, so we suspect metallicity to be even slightly higher 
(in the order of hundreths) than the obtained value $[\mathrm{Fe/H}]=-1.89$.

This value differs by an order of a few tenths from the published values: $[\mathrm{Fe/H}]=-2.5$ 
\citep{butler1982}, $[\mathrm{Fe/H}]=-2.22$ \citep{suntzeff1994}, $[\mathrm{Fe/H}]=-2.44$ \citep{fernley1998}. 
These values came from the $\Delta S$ approach \citep{preston1959}. In the case of the Blazhko star TV Boo, 
we need to measure $\Delta S$ in various Blazhko phases to obtain a meaningful value of $\Delta S$. 
The problem of previous studies is that $\Delta S$ was not determined at minimum light, as it is required. 
Furthermore it was determined only in in a few very close pulsation phases.

        \begin{figure}%[htbp]
            \begin{center}
            \includegraphics[scale=0.37]{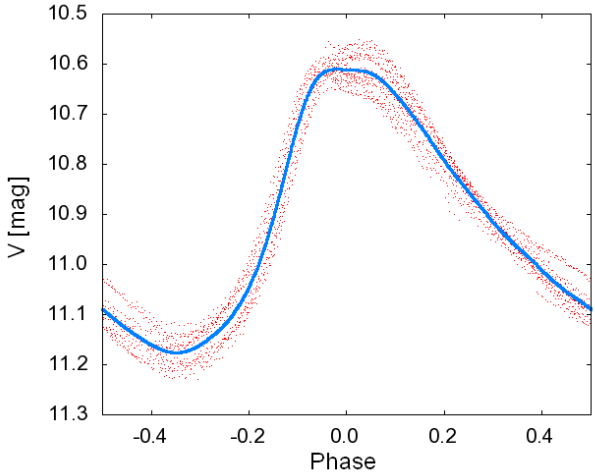}
            \caption{Phased MUO $V$ light curve together with our solution. The data are plotted as 
            points, the fit with continuous line.}
            \end{center}
        \end{figure}

Published values of $\Delta S$ vary between $8$ \citep{preston1959} and $13.1$ \citep{butler1975}. 
Transforming them via eq. 3 in \citet{suntzeff1994} we obtained a range of metallicities from $-1.67$ 
\citep{preston1959} to $-2.44$ \citep{butler1975}. Solving these ten light curves assuming one unique 
Blazhko cycle as shown in fig. 5, we obtained metallicities between $-1.68$ and $-2.0$, so the 
determination of metallicity strongly depends on the Blazhko phase.

\citet{pena2009} observed TV Bootis and determined basic physical parameters in the same way as we did. 
They proposed the metallicity to be $[\mathrm{Fe/H}]=-2.04$. Unfortunately, their measurements cover 
only one pulsation cycle.

The errors of physical parameters were calculated according to errors of calibrations given in papers 
they are taken from. All parameters (except metallicity) of TV Bootis in tab. 5 are almost the same as 
found in literature, especially with those listed in \citet{pena2009}. The value of the radius of star 
is only rough estimation determined by solving Stefan-Bolzmann equation for a black body.

The estimated mass of TV Boo is quite high compared to RRab type stars, but it is just in the middle of 
mass interval for RRc stars noted in \citet{simon1993}. The distance $r$ was determined using the distant 
modulus with extinction 0.024 mag in the $V$ band\footnote{taken from \\http://ned.ipac.caltech.edu/forms/calculator.html}. 
This value is in agreement with 1149 pc noted in \citet{liu1990}.

\section{Conclusions}

The Blazhko star TV Bootis was analysed using data gathered at MUO in 23 nights together with data 
from the SuperWASP survey. The data cover few years and contain a wide range of Blazhko phases. 
Therefore, we were able to determine new pulsation properties with the best precision ever and also 
we were able to obtain meaningful values of physical parameters.

Frequency analysis revealed symmetric structures located in the vicinity of the basic pulsation 
frequency and its harmonics with two modulation peaks. The star with the basic pulsation period 
a little bit shorter than 8 hours shows light curve modulation with a period of 9.7374~d. This is one 
of the shortest Blazhko period known for RR Lyrae type stars. Frequency analysis also unveiled another 
modulation period with a value of 21.5~d, which causes changes in Blazhko effect itself. In addition 
to this period, there are some indications for longer light curve changes in the order of years. This behavior 
should be confirmed or disproved by further observations. No signs of the period doubling phenomenon or 
a strange frequency in the range 1.58--1.63 $f_{0}$, revealed in some RRc stars, were detected.

We found TV Bootis to be a low-metallicity star with $[\mathrm{Fe/H}]=-1.89$. This value is slightly higher 
than metallicities found in literature (from -2.5 to -2.04). These values come from only few measurements 
based mainly on spectroscopic observations with short timespan, carried out in one unique Blazhko phase. 
The lack of observations in different Blazhko phases and low number of observations is a crucial deficiency 
in many studies of Blazhko stars. The discrepancies in metallicities could be resolved by a new spectroscopic 
measurements made in many Blazhko phases with sufficient timespan.

\section*{Acknowledgments}

We would like to thank Zden\v ek Mikul\'a\v sek, S. N. de Villiers and our referee for useful 
comments and suggestions. We acknowledge the WASP consortium which comprises of the University of 
Cambridge, Keele University, University of Leicester, The Open University, The Queen's University 
Belfast, St. Andrews University and the Isaac Newton Group. Funding for WASP comes from the consortium 
universities and from the UK's Science and Technology Facilities Council. This work was supported by 
GACR project GD205/08/H005, MU MUNI/A/0968/2009 and GAP 209/12/0217.

\bsp

\label{lastpage}

\end{document}